\documentclass[amsmath, amssymb, superscriptaddress, nofootinbib]{revtex4-2}

\usepackage{graphicx}
\usepackage{natbib}
\usepackage{amsmath,amssymb}
\usepackage{caption}
\usepackage{float}
\usepackage{dcolumn}
\usepackage{bm}
\usepackage[colorlinks=true,citecolor=blue,linkcolor=blue,urlcolor=blue]
{hyperref}
\usepackage{subfig}
\usepackage{braket}
\usepackage{mathrsfs}
\usepackage{scalerel}[2016/12/29]
\usepackage{enumerate}
\usepackage{dsfont}
\usepackage{mathtools}

\usepackage{verbatim} 
\usepackage{scrextend} 
\usepackage{float}
\usepackage[utf8]{inputenc} 

\begin{document}
	
	\title{Properties of black hole exterior fermions in spinfoam}
	
	
	\author{Ayanendu Dutta}
	\email{ayanendudutta@gmail.com}
	\affiliation{Department of Physics, Jadavpur University, Kolkata-700032, India}
	

	
\begin{abstract}
We discuss fermion tunneling from the black hole exterior to white hole exterior geometry using the spin foam technique in background-independent Loop Quantum Gravity. The fermion transition time is equal, longer, or shorter than the black hole tunneling time, depending on the Euclidean dihedral angle. For a locally pinched negative curvature, fermions accumulate in the black hole exterior region where the black hole has already emerged from the white hole. Conversely, in locally stretched positive curvature, fermions reach the white hole exterior prior to the black hole, indicating an Einstein-Rosen bridge-like geometry in black hole-to-white hole bounce.
\end{abstract}
\maketitle

\section{Introduction}
	
Recent development of black hole to white hole bouncing scenario offers a new fundamental method to represent the fate of a black hole when they reach the `Planck star' stage \cite{Rovelli:2014cta}, according to quantum effects. One of the promising step in this context was put forward by Haggard and Rovelli \cite{Haggard:2014rza}, where they have showed that, a violation of classical field equations within a finite spacetime is ``sufficient to allow'' a black hole ``trapped region'' tunnel into a white hole ``anti-trapped region''. In an alternative terminology, quantum fluctuation in the metric triggers geometry transition and the matter trapped inside a black hole can emerge from white hole. The results describe a metric that is allowed by Einstein's equations almost everywhere except the quantum tunneling region. 
According to dimensional arguments, small quantum effects accumulate to initiate the tunneling of a collapsed object of mass $ m $, after a time $ \tau \sim m^2 $ in Planck scale. This is sufficiently shorter than the subdominant Hawking evaporation time $ \tau_{H} \sim m^3 $. For the references of relevant astrophysical phenomena, see \cite{Christodoulou:2016vny}.

A key ingredient in this direction of research is to compute the characteristic time scales of the geometry transition using Loop Quantum Gravity (LQG) amplitudes and spinfoam. Spinfoams \cite{Baez:1997zt} combine concepts from topological quantum field theories, covariant lattice quantization, and the canonical quantization program of LQG. These models are characterized by spin-state-sum, which specify the regularized partition function through the quantization of geometrical shapes. From the Ponzano-Regge model \cite{Ponzano1968,Regge:1961px}, to the recent EPRL amplitudes \cite{Engle:2007wy,Freidel:2007py}, these models have undergone substantial advancements. Various studies \cite{Christodoulou:2016vny,Christodoulou:2018ryl} have already successfully estimated the lifetime $ \tau $ and the crossing time (tunneling time) $ T_c $ of a black hole, using geometry transition through Lorentzian EPRL model. On the other hand, the dynamical coupling of fermions to quantum gravity has a difficulty to address previously, until the \textit{``Spinfoam fermion''} amplitude has been successfully analyzed \cite{Bianchi:2010bn,Han:2011as}. See \cite{Rovelli:2014ssa,Perez:2012wv} for detailed reviews on these grounds.

The definition of the stable vacuum exterior geometry of the Haggard-Rovelli (HR) metric denotes the usual Schwarzschild exterior spacetime. However, if coupled fermions are present in this exterior region, their transition is still a missing ingredient in the theory. So, for the further advancement of the model, the fate of the particles, which could emerge from the quantum fluctuation, should be addressed alongside. Hence, the study, intended to be proposed here, has twofold motivation. First, consider that the vacuum fluctuation produces enough fermions in the black hole exterior region \cite{Kuzmin:1998kk,Adorno:2021xvj}, which in turn, is coupled to gravity. Second, they subsequently tunnel through spin foams to be released in the white hole exterior. Nevertheless, the manifestation is simple. The technical advantage has streamlined the investigation into derivation of the transition of spinfoam fermions alone.

In conventional covariant LQG, the amplitude for the geometry transition is given by
\begin{eqnarray}
	W(m,T) = \braket{W|\Psi_{m,T}},
	\label{amp-eq}
\end{eqnarray}
where $ \bra{W} $ and $ \ket{\Psi_{m,T}} $ respectively denote the spinfoam amplitude and semiclassical coherent boundary state that peaked on the 3d geometry of the entire boundary. For this particular context, $ \ket{\Psi_{m,T}} $ corresponds to Thiemann's $ SL(2,\mathbb{C}) $ heat-kernel coherent states given by \cite{Thiemann:2000bw,Thiemann:2000ca,Thiemann:2000bx,Thiemann:2000by}. On the other hand, from the definition, the crossing time is readily written as
\begin{eqnarray}
	T_c \sim \frac{\int dT~T~|W(m,T)|^2 }{\int dT~|W(m,T)|^2},
	\label{cross-time}
\end{eqnarray}
where $ p\sim |W(m,T_c)|^2 $ is the tunneling probability, and the calculations are provided in geometrical units, i.e. $ G=c=1 $. The propagators here are defined by the transition amplitude of the quantum transition region. The amplitude certainly depends on $ m $ and $ T $, since the external HR geometry depends on them. Here, the mass $ m $ is the only fixed quantity in this estimation. By definition, the amplitude also depends on the interior boundary choice, but, in particular, the time scales must be independent from such choices.

From the estimation of time scales of black hole transition, the transition amplitude is provided by contracting the EPRL spin foam amplitude with the coherent boundary state from the heat-kernel. For the derivation of crossing time $ T_c $ and the lifetime $ \tau $ of a black hole, the form of EPRL amplitude for physically pertinent Lorentzian case in 3d is realized as
\begin{eqnarray}
	W(h_\ell)= \int_{SU(2)} dh_{vf} \prod_{f} \delta(h_f) \prod_{v} A_v(h_{vf}).
	\label{eprl-amp}
\end{eqnarray}
This equation remains unchanged in 4d, since, 4d kinematic Hilbert space is equivalent in 3d. The vertex amplitude however, is written as \cite{Rovelli:2014ssa} \footnote{For the transition amplitude in Eq. \eqref{eprl-amp} and \eqref{vertex-eq}, refer to equations (5.72), (5.74), (5.77), (7.46), (7.47), (7.57) and (7.58) in Rovelli's book, ``Covariant Loop Quantum Gravity'' \cite{Rovelli:2014ssa}.}
\begin{eqnarray}
	A_v(h_{vf})=\sum_{j_f} \int_{SU(2)} dg'_{ve} \prod_{f} dj_f~Tr^{j_f} \bigg[ g_{e'v} g_{ve} h_{vf} \bigg].
	\label{vertex-eq}
\end{eqnarray}
After essential rearrangements, one may address the EPRL vertex transition amplitude after $ Y_{\gamma} $ mapping as \footnote{Refer to equations (2.109), (2.110), (3.1) and (3.2) in \cite{Marios:2018hzw}, and Eq. (7.48) in Rovelli's book \cite{Rovelli:2014ssa}.
\\	
Note that, without loss of generality, $ Tr^{j_f} $ in Eq. \eqref{vertex-eq-2} is sometimes written as $ Tr^{\gamma j_f,j_f} $.}
\begin{align}
	W(h_\ell) &= \int_{SU(2)} \mu(h_{vf}) \prod_{f} \delta(h_f) \prod_{v} A_v(h_{vf}, g_{ve}, j_f), \label{eprl-amp-2}
	\\
	A_v(h_{vf}, g_{ve}, j_f) &= \sum_{\{j_f\}} \int_{SL(2,\mathbb{C})} \mu(g_{ve}) \prod_{f} dj_f ~Tr^{j_f} \bigg[ \prod_{v\in f} Y^{\dagger}_{\gamma} g_{e'v} g_{ve} Y_{\gamma} h_{vf} \bigg].
	\label{vertex-eq-2}
\end{align}
where $ j_{f} $ is the spin-$ j $ representation of the quantum gravity vertex amplitude.

Consider now the minimal coupling of fermions to the covariant dynamics of Loop Quantum Gravity. The discretized action of fermions in interaction with gravity is given in the $ SL(2,\mathbb{C}) $ matrix $ g_e $, as
\begin{eqnarray}
	S=i \sum_{e} v_e \bar{\psi}_{s_e} g^{\dagger}_e g_e \psi_{t_e},
	\label{fer-action}
\end{eqnarray}
and the amplitude of each cycle as \cite{Bianchi:2010bn}
\begin{eqnarray}
	A_c = (-1)^{|c|}~Tr^{1/2}\bigg(\prod_{e \in c} \big(g_{e s_e} g^{-1}_{e t_e}\big)^{\epsilon_{ec}}\bigg),
	\label{fer-cyc-amp}
\end{eqnarray}
where $ \epsilon_{ec}=\pm 1 $, depending upon whether the edge and cycle orientation agrees, and, $ s_e $ and $ t_e $ are the source and target of edge $ e $.

The fermion fields are characterized by a chiral spinor $ \textbf{n} \in \mathbb{C}^2 $ in $ H^{1/2} $, with two complex components $ n^A $, $ A=0,1 $. The spinor $ \textbf{n}=(0,1) $ is the coherent state $ \ket{1/2,~1/2} $, since all other coherent states are obtained rotating $ \ket{1/2,~1/2} $ in the $ j=1/2 $ representation. For fermions, a coherent state can already be represented by the linear function 
\begin{eqnarray}
	f_\textbf{n}(\textbf{z}) \sim n^A z_A \sim \braket{\textbf{z}|\textbf{n}},
\end{eqnarray}
however, in $ j $-representation, (after normalization) this should express the coherent state of angular momentum $ j \vec{n} $ as
\begin{eqnarray}
	f^j_\textbf{n}(\textbf{z}) \sim \sqrt{\frac{2j+1}{\pi}} \braket{\textbf{z}|\textbf{n}}^{2j}.
\end{eqnarray}
The spin-$ j $ representation space $ H_j $ can be represented using $ 2j $ spinor indices. Additionally, another useful realization of the spin spaces $ H_j $ involves functions $ f(\textbf{z}) $ of spinors. Specifically, the finite-dimensional vector space $ H_j $ can be understood as the space of totally symmetric polynomial functions $ f(\textbf{z}) $ of degree $ 2j $.

More importantly, in the discretized fermion action, the 3-volume $ V_e $ is determined by the spins $ j_f $ and the coherent intertwiner $ i_e $. Specifically, the vertex amplitude $ A_v[j_f,i_e,g_{ve}]_{GR} $ can be expressed in terms of Livine-Speziale coherent intertwiners $ i_e=\| \vec{j}_f, \vec{n}_{ef} \rangle $, where the labels $ (\vec{j}_f, \vec{n}_{ef}) $ define the geometry of a tetrahedron (or polyhedron). These labels, corresponding to spins $ j_f $ and unit vectors $ \vec{n}_{ef} $, must satisfy the closure condition $ \sum_{f} j_f \vec{n}_{ef}=0 $. Clearly, for the coupling of fermions to spinfoam quantum gravity, the Dirac fermion part holds an extra edge amplitude $ A_e[\psi_{b(e)},\psi_{f(e)},g_{ve},j_f,i_e] = e^{i S_e} $, with the gravity part; where the fermion action $ S_F(\mathcal{K},j_f,i_e,g_{ve},\psi_v) $ in a 2-complex $ \mathcal{K} $, is invariant under $ SL(2,\mathbb{C}) $ gauge transformation $ G_{ve} \rightarrow \Lambda_v G_{ve}, ~ \psi_v \rightarrow \Lambda_v \psi_v $ and $ \bar{\psi}_v \rightarrow \bar{\psi}_v \Lambda^{-1}_v $ ($ \Lambda_v $ and $ G_{ve} $ are the $ SL(2,\mathbb{C}) $ elements $ \lambda_f $ and $ g_{ve} $ represented on the space of Dirac spinors). 

Moreover, the fermionic Fock space must be introduced here to accommodate the explanations of quantum field theory. Combining the fermionic Fock space with the kinematics of quantum gravity involves locating fermions on the nodes $ n $ of a graph, akin to Lattice gauge theory. Each node of the graph is associated with a copy of the Fock space $ F $. Consequently, the states of the gravity+fermion theory exist within the space $ (\bigotimes_\ell L_2[SU(2)]) \bigotimes (\bigotimes_n F) $, where $ \ell $ labels the links of the graph and $ n $ labels the nodes. These states are constrained by the gauge action of $ SU(2) $ at each node, and can be denoted as $ \Psi_{h_\ell, \psi_n} $.

The spin networks, forming a basis of these states, extend the concept of pure gravity spin networks. It is advantageous to select an intertwiner basis at each node $ n $ that diagonalizes the volume of that node, labeled with the volume eigenvalue $ v_n $, denoted as $ \ket{j_n, v_n} $. In the presence of fermions, spin networks carry an additional quantum number $ c_n $ at each node, corresponding to the basis $ \ket{c} $ in the Fock space at that node, denoted as $ \ket{j_n, v_n, c_n} $. At a $ v $-valent node $ n $, bounded by links with spins $ j_1,\cdots,j_n $, the intertwiner $ v_n $ forms an invariant tensor in the tensor product of the $ v $ representations $ j_1,\cdots,j_n $ if $ c_n = \emptyset $ or $ c_n=2 $. However, if $ c_n=\pm $, then $ v_n $ acts as an invariant tensor in the tensor product of the $ v+1 $ representations $ j_1,\cdots,j_v,\frac12 $. In this scenario, the intertwiner couples the spinor to the gravitational magnetic indices.

The Fock notation is used here, to conveniently represent states within this framework. The state where all nodes are in the Fock vacuum, is denoted as $ \ket{j_n, v_n} $, indicating that no particles are present at any node, i.e., $ \ket{j_n, v_n, c_n=\emptyset} $. A state with all nodes in the vacuum except for a single node $ n $ containing a particle with spin $ a $ is denoted as $ \ket{j_n, v_n, (n,a)} $. For a state with $ N $ particles distributed among nodes $ n_1,\cdots,n_N $ with spins $ \pm n $, it is represented as $ \ket{j_n, v_n, (n_i, \pm)} $ with $ i=1,\cdots,N $. Two-particle singlets, where two fermions with opposite spins occupy the same node, are denoted as $ \ket{j_n, v_n, (n,+), (n,-)} $. This notation encapsulates the boundary kinematical state space of the theory. For a detailed discussion, readers are referred to \cite{Rovelli:2014ssa, Han:2011as, Bianchi:2010bn}.

Hence, it is convenient to denote the fermion spins in the $ j $-representation, namely by $ j_{\bar{f}} $. Note that to avoid risking the loss of generality, a different spin-$ j $ representation is considered for fermions, as compared to the gravity. Now, in this notation, Eq. \eqref{fer-cyc-amp} can be rewritten in an alternative form as
\begin{equation}
	A_c = (-1)^{|c|} Tr^{j_{\bar{f}}} \left[g^{\dagger}_{e_1 v_1} g_{e_1 v_2} \cdots g^{\dagger}_{e_n v_n} g_{e_n v_1} \right] .
	\label{fer-amp-alternative}
\end{equation}
Note that, $ (v_1,e_1,v_2,e_2,\cdots, v_n,e_n) $ is the sequence of vertices and oriented edges that are crossed by the cycle $ c $.


Now, functions defined on the $ SL(2, \mathbb{C}) $ can be expressed as expansions in terms of irreducible unitary representations of $ SL(2, \mathbb{C}) $, characterized by a positive real number $ p $ and a non-negative half-integer $ k $ in the $ V^{(p,k)} $ representation space. The $ Y_\gamma $ map introduces the following transformations: $ H_j \rightarrow V^{(p=\gamma j, k=j)}, ~ L_2[SU(2)] \rightarrow F[SL(2,\mathbb{C})], ~ \ket{j,m} \rightarrow \ket{\gamma j,j,j,m} $. The physically realizable states in quantum theory of gravity are $ SU(2) $ spin networks, or their mapping under $ Y_\gamma $. $ Y_\gamma $ map relies on the Einstein-Hilbert action and encodes how $ SU(2) $ states transform under $ SL(2,\mathbb{C}) $ transformations in the theory.

A convenient way to realize the representation space of $ V^{(p,k)} $ is through functions of spinors, where $ \textbf{z} \in \mathbb{C}^2 $. This representation $ (p,k) $ is defined on the space of homogeneous functions of spinors that possess the following property
\begin{eqnarray}
	f(\lambda \textbf{z}) = \lambda^{-1+ip+k} \bar{\lambda}^{-1+ip-k} f(\textbf{z}) ,
\end{eqnarray}
which can also be modified to
\begin{eqnarray}
	f^{j}_{m}(\textbf{z}) = \braket{\textbf{z}|p,k,j,m} = \sqrt{\frac{2j+1}{\pi}} \braket{\textbf{z}|\textbf{z}}^{ip-1-j} D^{j}_{mk}(g(\textbf{z})) ,
\end{eqnarray}
in the Wigner-$ D $ form. The form of the $ Y_\gamma $ map when it acts on spinor states is notably straightforward. This expression reveals how the coherent state $ \ket{j,\vec{n}} $ appears after it undergoes mapping by $ Y_\gamma $ into the suitable $ SL(2,\mathbb{C}) $ representation.

Note that a fermion essentially denotes an extra face of spin $ 1/2 $, acting non-locally on the 2-complex \cite{Bianchi:2010bn}. The extra face is said to be the bulk face, bounded by the sequence of ertices and edges. For fermions, in every half edge within the bulk, a group element $ g_{ve} \in SL(2,\mathbb{C}) $ is assigned, where conventionally $ g_{ve}=g^{-1}_{ev} $. When an edge originally form vertex $ v $ terminates at node $ n $, it remains unsplit, and the associated group element is denoted as $ g_{vn} \in SL(2,\mathbb{C}) $. Links are characterized by $ SU(2) $ group element $ h_\ell $, and all faces, are labeled by half-integer spins $ j_{\bar{f}}>0 $. The orientation of faces determines the distinction between ingoing and outgoing (source and target) group elements. Specifically, an element of the form $ g{ev} $ resides on the half edge $ e $ and entering vertex $ v $ is termed ingoing, while $ g_{ve'} $ on the half edge $ e' $ exiting vertex $ v $ is termed outgoing.

The amplitude $ W $ can be expressed as a product of bulk face amplitudes, each associated with every face of the 2-complex. For bulk faces, the face amplitude $ A_f $ is constructed as follows: At each vertex $ v $, the product of ingoing group element $ g_{ev} $ is formed, and outgoing element $ g_{ve'} $, denoted as $ g_{ev} g_{ve'} $. Each such product is then multiplied from the left by $ Y_\gamma^\dagger $ and from the right by $ Y_\gamma $, resulting in a product of the form, $ Y_\gamma^\dagger g_{ev} g_{ve'} Y_\gamma $, at every vertex. The terms combine as the face is traversed, and the face amplitude is defined accordingly. Thus, the coupling of fermions to gravity, the fermion assigns an bulk face amplitude $ A_f(g_{ev}) $ with the quantum gravity vertex amplitude. 

The negative sign $ (-1) $ in Eq. \eqref{fer-amp-alternative} originates from the orientation of fermion cycles with the corresponding edge \( e \). By considering fermion arrow lines directed oppositely to the edge \( e \), this can be approximated \cite{Han:2011as}. Hence, by dropping the cycle index $ c $ to alleviate the notations, the amplitude is given by
\begin{align}
	A_f(g_{ev}) = \sum_{j_{\bar{f}}} Tr^{j_{\bar{f}}} \left[ Y_\gamma^\dagger g^{\dagger}_{ve} g_{ve'} Y_\gamma Y_\gamma^\dagger g^{\dagger}_{v'e'} g_{v'e''} Y_\gamma \cdots Y_\gamma^\dagger g^{\dagger}_{v^{(n)} e^{(n)}} g_{v^{(n)} e} Y_\gamma \right] = \sum_{j_{\bar{f}}} Tr^{j_{\bar{f}}} \left[ Y_\gamma^\dagger g_{ev} g_{ve'} Y_\gamma \right].
	\label{fermion-amp}
\end{align}
This is the desired $ Y_\gamma $ representation of the fermionic amplitude with $ SL(2,\mathbb{C}) $ holonomy $ g_{ev} $ of the spin connection.
\\
Therefore, the final ansatz of the spinfoam amplitude for the dynamics of gravity-fermion coupling is defined as \cite{Bianchi:2010bn}
\begin{eqnarray}
	W = \sum_{j_f, j_{\bar{f}} } \int_{SL(2,\mathbb{C})} dg'_{ve} \int_{SU(2)} dh_{vf} \prod_{f} \delta(h_f) ~ dj_f \underbrace{Tr^{j_f} \Bigg[\prod_{v \in f} g_{e'v} g_{ve} h_{vf} \Bigg]}_{gravity~part} \times \underbrace{Tr^{j_{\bar{f}}} \Bigg[\prod_{v \in f} g_{ve'} g_{ev} \Bigg]}_{fermion~part}.
	\label{fer-amp-final}
\end{eqnarray}
or consequently
\begin{eqnarray}
	W=W(h_\ell)= \int_{SU(2)} dh_{vf} \prod_{f} \delta(h_f) \prod_{v} A_v \prod_{f} A_f \, ,
	\label{eprl-fer-amp}
\end{eqnarray}
where $ A_v $ and $ A_f $ are given by equations \eqref{vertex-eq-2} and \eqref{fermion-amp} respectively. Note that the $ SL(2,\mathbb{C}) $ matrices $ g_{ev} $ share the same geometric interpretation as operators transporting in parallel from the edge to the vertex. More significantly, the analysis of the vertex amplitude demonstrates that the saddle point approximation of the integral depends on the value of $ g_{ev} $. This value rotates the Lorentz frame of the 4-cell into a Lorentz frame at the 3-cell, aligning the time direction with the normal of the 3-cell. Therefore, as one moves away from the Planck scale, these group elements precisely adopt the value required to produce the fermion action \cite{Rovelli:2014ssa}.

%
%

\section{Han-Krajewski path integral formulation}

Using the $ V^{\gamma j, j} $ of $ SL(2,\mathbb{C}) $ representation through $ Y_\gamma $ map, the trace of the vertex amplitude of gravity sector \eqref{vertex-eq-2} can be written as
\begin{equation}
	Tr^{j_f}\left[ \prod_{v \in f} Y^\dagger_\gamma g_{e'v} g_{ve} Y_\gamma h_{vf} \right] = \sum_{\{m\}} \prod_{v \in f} {}_\gamma\braket{j_f~m_{e'}| g^{-1}_{ve'} g_{ve}| j_f~\tilde{m}_e}_\gamma \braket{j_f~\tilde{m}_e|h_{vf}|j_f~m_{e}}.
\end{equation}
After eliminating $ h_{vf} $, the EPRL model reduces to
\begin{equation}
	A_v(j_f) = \sum_{\{m\}} \prod_{v \in f} {}_\gamma\braket{j_f~m_{e'}| g^{-1}_{ve'} g_{ve}| j_f~m_e}_\gamma \, .
	\label{path-integral-1}
\end{equation}
Now, to express this EPRL amplitude to the desired form, the Han-Krajewski path integral approach \cite{Han:2013gna} is employed. \footnote{Readers may refer to chapter 8.3 of Rovelli's book \cite{Rovelli:2014ssa}, and \cite{Marios:2018hzw} for extended calculations} In the $ V^{\gamma j, j} $ space, the explicit homogeneous spinor basis $ z $ is realized as
\begin{equation}
	\braket{z|j~m}_\gamma = f^j_m(z) = \sqrt{\frac{2j+1}{\pi}} \braket{z|z}^{j(i\gamma-1)-1} D^j_{mj}(H(z)) ,
\end{equation}
where the function $ H(z) $ is given by
\begin{equation}
	H(z) = 
	\begin{pmatrix}
		z_0 & -\bar{z}_1 \\
		z_1 & \bar{z}_0
	\end{pmatrix}
\end{equation}
in the spinor basis, and $ D^j_{mj}(H(z)) $ is the Wigner-D matrix for the function $ H(z) $. The transformation of $ \braket{z|j~m}_\gamma = f^j_m(z) $ under the group $ SL(2,\mathbb{C}) $ is performed through the transpose matrix action in the fundamental representation of $ SL(2,\mathbb{C}) $, treating the spinor argument
\begin{equation}
	g \triangleright f^j_m(z) = f^j_m(g^T z) .
\end{equation}
When considering $ g^{-1}_{ve'} g_{ve} $ as a unified $ SL(2,\mathbb{C}) $ element operating on the right via the transpose action as \cite{Barrett:2009mw}
\begin{equation}
	{}_\gamma\braket{j_f~m_{e'}| g^{-1}_{ve'} g_{ve}| j_f~m_e}_\gamma = \int_{\mathbb{CP}^1} d\mu(z) \overline{f^{j_f}_{m_{e'}}(z_{vf})} f^{j_f}_{m_{e}}\left(g^T_{ve}(g^{-1}_{ve'})^T z_{vf}\right) ,
	\label{path-integral-2}
\end{equation}
where $ \mu(z) $ is the $ SL(2,\mathbb{C}) $ invariant integration measure on $ \mathbb{C}^2 $, conveniently written as
\begin{equation}
	\mu(z)= \frac{i}{2} (z_0 dz_1 - z_1 dz_0) \wedge (\bar{z}_0 d\bar{z}_1 - \bar{z}_1 d\bar{z}_0).
\end{equation}
Under the $ SL(2,\mathbb{C}) $ invariant measure $ \mu(z) $, one may modify the spinor variable as $ z_{vf} \rightarrow (g^{-1}_{ve'})^{T} z_{vf} $, which rearranges Eq. \eqref{path-integral-2} as
\begin{equation}
	{}_\gamma\braket{j_f~m_{e'}| g^{-1}_{ve'} g_{ve}| j_f~m_e}_\gamma = \int_{\mathbb{CP}^1} d\mu(z) \overline{f^{j_f}_{m_{e'}}(g^T_{ve'} z_{vf})} f^{j_f}_{m_{e}}\left(g^T_{ve} z_{vf}\right) .
\end{equation}
Since $ \mu(g_{ve}) $ is a left and right invariant Haar measure, one may write $ g^T_{ve} \rightarrow g^{\dagger}_{ve} $, which may also follows \cite{Han:2013gna,Han:2011re}
\begin{equation}
	Z_{vef} = g^{\dagger}_{ve} z_{vf}, \qquad Z_{ve'f} = g^{\dagger}_{ve'} z_{vf} .
\end{equation}
Therefore, the vertex amplitude takes the form
\begin{equation}
	A_v(j_f,\{g_f\}) = \int \mu(g_{ve}) \sum_{\{m\}} \prod_{v\in f} \int_{\mathbb{CP}^1} d\mu(z) \overline{f^{j_f}_{m_{e'}}(Z_{ve'f})} f^{j_f}_{m_{e}}(Z_{vef}) .
\end{equation}
In Eq. \eqref{path-integral-1}, the summation can be aggregated over the magnetic index $ m_e $ for the two half-edges $ ev $ and $ e'v $, and merge the two occurrences of $ f^{j_f}_{m_{e}} $. As a result, the amplitude is expressed as 
\begin{equation}
	A_v(j_f) = \int \mu(g_{ve}) \int_{\mathbb{CP}^1} d\mu(z) \prod_{e\in f} \sum_{m_e} f^{j_f}_{m_{e}}(Z_{vef}) \overline{f^{j_f}_{m_{e}}(Z_{v'ef})} .
\end{equation}
Note that, for two spinor $ Z $ and $ W $, the Wigner-D can be written as \cite{Marios:2018hzw}
\begin{equation}
	\sum_m D^{j}_{mj}(H(W)) D^{j}_{mj}(H(Z)) = \braket{Z|W}^{2j} ,
\end{equation}
thereby, applying $ Z\rightarrow Z_{v'ef} $ and $ W\rightarrow Z_{vef} $:
\begin{equation}
	\sum_{m_e} f^{j_f}_{m_{e}}(Z_{vef}) \overline{f^{j_f}_{m_{e}}(Z_{v'ef})} = \frac{2j+1}{\pi} \braket{Z_{vef}|Z_{vef}}^{j(i\gamma-1)-1} \braket{Z_{v'ef}|Z_{v'ef}}^{j(-i\gamma-1)-1} \braket{Z_{v'ef}|Z_{vef}}^{2j} .
\end{equation}
Another important observation is that, aside from the factor $ \braket{Z_{vef}|Z_{vef}}^{-1} \braket{Z_{v'ef}|Z_{v'ef}}^{-1} $, which can be incorporated into a redefinition of the measure $ \mu(z) $, all other terms are raised to the power of $ j $. This significant result enables the application of the stationary phase approximation. Consequently, disregarding constants that can be absorbed into the overall normalization of the amplitude and terms that can be absorbed into integration measures, such as $ d_j=(2j+1) $ being absorbed into $ \mu(j_f) $, one may demonstrate that \cite{Christodoulou:2018ryl}
\begin{equation}
	A_v(j_f) = \int \mu(z_{vf}) ~e^{j_f F_f(\{g_f\},\{z_f\})} ,
\end{equation}
where
\begin{align}
	&\sum_{\{m\}} \prod_{v\in f} {}_\gamma\braket{j_f~m_{e'}| g^{-1}_{ve'} g_{ve}| j_f~m_e}_\gamma = e^{j_f F_f(\{g_f\},\{z_f\})} ,
	\\
	F_f\big( \{g_f\},\{z_f\} \big) &\equiv \sum_{e \in f} \log{\frac{\braket{Z_{v'ef}|Z_{vef}}^2}{\braket{Z_{vef}|Z_{vef}} \braket{Z_{v'ef}|Z_{v'ef}}}}
	+i \gamma \log{\frac{\braket{Z_{vef}|Z_{vef}}}{\braket{Z_{v'ef}|Z_{v'ef}}}} .
\end{align}
Analogously, for the fermionic face amplitude:
\begin{equation}
	A_f(j_{\bar{f}}) = \sum_{j_{\bar{f}}} Tr^{j_{\bar{f}}} \left[\prod_{v \in f} Y^\dagger_\gamma g_{ve'} g_{ev} Y_\gamma \right] = \sum_{\{m\}} \prod_{v \in f} {}_\gamma\braket{j_{\bar{f}}~m_{e'}| g^{-1}_{e'v} g_{ev}| j_{\bar{f}}~m_e}_\gamma \, ,
\end{equation}
which is very much equivalent to \eqref{path-integral-1}. Therefore, proceeding in the same line of approach as before, one may readily obtain
\begin{equation}
	\sum_{\{m\}} \prod_{v\in f} {}_\gamma\braket{j_{\bar{f}}~m_{e'}| g^{-1}_{e'v} g_{ev}| j_{\bar{f}}~m_e}_\gamma = e^{j_{\bar{f}} G_f(\{g_f\},\{z_f\})} ,
\end{equation}
leading to the final expression of the spinfoam amplitude for gravity-fermion coupling as follows:
\begin{eqnarray}
	W= \sum_{ \{j_f,j_{\bar{f}}\} } \mu(j_f) \int \mu(z_{vf}) ~e^{j_f F_f(\{g_f\},\{z_f\})}
	\times e^{j_{\bar{f}} G_f(\{g_f\},\{z_f\})}
	\label{grav-fer-exp}
\end{eqnarray}
The notation $ \{g_f\},\{z_f\} $ indicates a dependence on $ g_{ve} $ and $ z_{vf} $ with $ v \in f $. To be concise, one may use $ F_f(g,z) $ (or, $ G_f(g,z) $) instead of $ F_f(\{g_f\},\{z_f\}) $ (or, $ G_f(\{g_f\},\{z_f\}) $) in the following discussions.

%

\section{The spin-sum}

The boundary states under consideration here, are Thiemann's Heat-Kernel boundary state \cite{Thiemann:2000bw,Thiemann:2000ca,Thiemann:2000bx,Thiemann:2000by}, as described in the twisted-geometry parametrization \cite{Freidel:2010aq,Freidel:2010bw}. When represented in this manner, these states are also referred to as coherent spin-networks or extrinsic coherent states \cite{Rovelli:2014ssa,Bianchi:2009ky}. They belong to the truncated boundary Hilbert space $ H_\Gamma = L^2[SU(2)^L/SU(2)^N] $ and are labeled by data $ H_\ell $ derived from the discrete phase space $ P_\Gamma = {\times}_\ell ~ T^* SU(2)_\ell \approx {\times}_\ell ~ (\mathbb{R}^+_\ell \times S^1_\ell \times S^2_\ell \times S^3_\ell) $ of the twisted geometry, where $ L $ and $ N $ are respectively the number of links $ \ell $ and number of nodes $ n $ of the boundary graph $ \Gamma $. The heat-kernel boundary state is defined as
\begin{eqnarray}
	\Psi_{\Gamma;\, \omega_\ell, \zeta_\ell, \textbf{k}_{\ell n}}^t(h_\ell) = \sum_{\{j_\ell \}} \left( \prod_\ell d_{j_\ell} e^{-\left(j_\ell - \omega_\ell \right)^2 t \,+\, i \gamma j_\ell \, \zeta_\ell } \right) \psi_{\Gamma;\, \textbf{k}_{\ell n} }(j_\ell; h_\ell),
	\label{hk-state}
\end{eqnarray}
where $ \gamma $ is the Immirzi parameter, which is directly related to the smallest non zero quanta of area, and $ h_\ell \in SU(2),~d_j \equiv 2j+1 $. The boundary states under consideration are the gauge variant counterparts of coherent spin network states. These states exhibit a semiclassical nature, and are peaked on both the intrinsic and extrinsic geometry of a discretized boundary $ \mathcal{B} $. The states denoted as $ \Psi_{\Gamma;\, \omega_\ell, \zeta_\ell, \textbf{k}_{\ell n}}^t(h_\ell) $ are Gaussian superposition of the coherent states $ \psi_{\Gamma;\, \textbf{k}_{\ell n} }(j_\ell; h_\ell) $, which, in turn, are focused on the intrinsic geometry of the triangulation of $ \mathcal{B} $. It can be represented by the Wigner D-matrices form, as given by
\begin{eqnarray}
	\psi_{\Gamma,  \textbf{k}_{\ell n}}(j_\ell; h_\ell) = \prod_\ell \sum_{m_s m_t} D^{j_\ell}_{m_s j_\ell}(k_{s(\ell)}^\dagger) \; D^{j_\ell}_{m_t j_\ell}(k_{t(\ell)})  D^{j_\ell}_{m_s m_t}(h_\ell),
	\label{wignerD}
\end{eqnarray}
where, to represent corresponding 3d normals of the space, $ SU(2) $ element $ k $ is chosen precisely. The states $ \Psi_{\Gamma;\, \omega_\ell, \zeta_\ell, \textbf{k}_{\ell n}}^t(h_\ell) $ denote semiclassical states within the truncated kinematical state space. The gauge-invariant version, imposing $ SU(2) $ gauge invariance at each node of $ \Gamma $, was methodically introduced in \cite{Bianchi:2009ky}. It was demonstrated that these states correspond to the large spin limit of Thiemann's $ SL(2,\mathbb{C}) $ heat-kernel states, utilizing the twisted geometry parametrization.

The transition amplitude can now be calculated by contracting the final spin foam amplitude \eqref{grav-fer-exp} with the heat-kernel state \eqref{hk-state}. In the holonomy representation, the contraction is carried out by integrating across the boundary $ SU(2) $ elements $ h_\ell $. The resulting transition amplitude is then expressed as
\begin{eqnarray}
	W(\omega_\ell,\zeta_\ell,\textbf{k}_{\ell n}, t) = W^t_\tau(H_\ell) = \mathcal{N} \sum_{\{j_\ell \} \in D^n_\omega}  \mu_j \bigg( e^{-\left(j_\ell - \omega_\ell \right)^2 t \,+\, i \gamma j_\ell \, \zeta_\ell } \bigg)  \int_{D_{g,\textbf{z}}} d \mu_{g,\Omega} \bigg(e^{j_\ell\; F_\ell(g,\textbf{z};\textbf{k}_{\ell n})} \times e^{j_{\ell_f}\; G_\ell(g,\textbf{z};\textbf{k}_{\ell n})} \bigg) ,
	\label{final-tran-amp}
\end{eqnarray}
where $ j_{\ell_f} $ is the spins corresponding to fermions. The notation $ \int_{D_{g,\textbf{z}}} d \mu_{g,\Omega} $ is used to encompass all $ SL(2,\mathbb{C}) $ and $ CP^1 $ integrals, whereas the notation
\begin{eqnarray}
	\mu_j := \bigg(\prod_{f\in \Gamma} \prod_{e\in f} dj_\ell \bigg) \bigg(\prod_{\ell \in \Gamma} d^4_{j_\ell}\bigg),
	\label{sum-measure}
\end{eqnarray}
represents the summation measure. Any irrelevant factors have been absorbed into the normalization factor $ \mathcal{N} $. An interesting point here is that $ \mu_j $ particularly depends only on the geometry. Thus, the fermion spin factor $ j_{\ell_f} $ however, appears in the amplitude; it sheds no effect on the summation measure. the summation across the boundary spin is carried out over the scale
\begin{eqnarray}
	D^{\bar{n}}_{\omega} := \underset{\ell}{\times} \left\{\biggl\lfloor{\omega_\ell - \frac{\bar{n}}{\sqrt{2t}}}\biggr\rfloor, \biggl\lfloor{\omega_\ell + \frac{\bar{n}}{\sqrt{2t}}}\biggr\rfloor \right\},
	\label{boundary-scale}
\end{eqnarray}
where $ 0< \bar{n} \in \mathbb{N} $. In this context, $ \lfloor x \rfloor $ represents the floor function, which is defined as the largest half-integer number $ \le x $. The limitation to the summation domain $ D^{\bar{n}}_{\omega} $ enforces the truncation of the spin-sum. The Gaussian weight factor $ \tilde{G}_\ell [j_\ell,H_\ell]:= i\gamma j_\ell \zeta_\ell - (j_\ell -\omega_\ell(\eta_\ell,t_\ell))^2 t_\ell $ control the spin-sums, with $ \bar{n} $ serving as a cutoff. The cutoff parameter $ \bar{n} $ measures the quantity of standard deviations $ \sigma=\frac{1}{\sqrt{t}} $ from the peak $ \omega_\ell $.

Note that, the holomorphic amplitude in \eqref{final-tran-amp} cannot directly be considered as independent. Specifically, the term in the equation becomes zero if any of the intertwiner spaces linked to the nodes of the 2-complex has dimension zero. To handle the sums independently and interchange them with the integrals, it is necessary to limit the spin sums to configurations where the intertwiner space is consistently non-trivial. Therefore, assuming the nodes of the 2-complex to be four-valent, a set is introduced as 
\begin{eqnarray}
	D_{\Gamma} = \bigg\{\{j_\ell\} \bigg\vert \min\left(j_1+j_2, j_3+j_4\right) \max\left(\vert j_1 - j_2\vert, \vert j_3 - j_4\vert\right) + 1 > 0 \quad\forall n \in \Gamma\bigg\},
	\label{spin-set}
\end{eqnarray}
which is the set of spin configurations $ \{ j_{\ell} \} $, such that the intertwiner spaces across the complete boundary graph $ \Gamma $ are non-trivial. So, the cutoff parameter considers $ \{ j_{\ell} \} \in D^{\bar{n}}_\omega \subseteq D_\Gamma $.

Now, for the next steps, it is necessary to separate the boundary spins of gravity ($ j_\ell $) and the fermion sector ($ j_{\ell_f} $) into fluctuations and fixed background contributions, as
\begin{align}
	j_\ell &= \lambda \delta_\ell + a_\ell, \qquad \omega_\ell \equiv \lambda \delta_\ell; \qquad\text{for geometry sector} 
	\nonumber \\
	j_{\ell_f} &= \lambda \delta_{\ell_f} + a_\ell, \qquad  \omega_{\ell_f} \equiv \lambda \delta_{\ell_f}; \qquad\text{for fermion sector} 
	\nonumber \\
	a_\ell &\in \bigg\{-\biggl\lfloor{\frac{\bar{n}}{\sqrt{2t}}}\biggr\rfloor, \biggl\lfloor{\frac{\bar{n}}{\sqrt{2t}}}\biggr\rfloor \bigg\} \quad \forall\ell\in\Gamma,
	\label{JDecomposition}
\end{align}
where $ \lambda \delta_\ell $ and $ \lambda \delta_{\ell_f} $ are fixed background contributions for geometry and fermion sectors respectively. $ a_\ell $ is the fluctuation, which is supposed to be the same for both the geometry and fermion sectors. This decomposition states that $ \delta_\ell $ and $ \delta_{\ell_f} $ are supposed to be of the order unit in $ \lambda $ with $ \lambda \gg 1 $. For the relation of $ t $ with $ \lambda \delta_\ell $ and $ \lambda \delta_{\ell_f} $ within the semiclassicality condition, refer to \cite{Rovelli:2014ssa,Christodoulou:2023psv}, which states that the coherent states are constructed by combining intrinsic coherent states, each centered around a triangulation of a spacelike hypersurface, into a superposition.

Thus, following the decomposition \eqref{JDecomposition}, Eq. \eqref{sum-measure} reads
\begin{align}
	\mu_j 
	&= \left(\prod_{f\in\Gamma}\prod_{e\in f} (2 j_\ell+1) \right)\left(\prod_{\ell\in\Gamma}(2 j_\ell+1)^4\right)
	\approx \left(\prod_{f\in\Gamma}\prod_{e\in f} 2 j_\ell \right)\left(\prod_{\ell\in\Gamma}(2 j_\ell)^4\right) 
	\nonumber \\
	&= 2^{M_\mathcal{C}}\left(\prod_{f\in\Gamma}\prod_{e \in f} (\lambda \delta_\ell+a_\ell) \right)\left(\prod_{\ell\in\Gamma}(\lambda \delta_\ell+a_\ell)^4\right)
	=\left(2 \lambda \delta_\ell\right)^{M_\mathcal{C}}\left(1+\mathcal{O}\left(\frac{a_\ell}{\lambda \delta_\ell}\right)\right) .
	\label{mu_j}
\end{align}
Referring to the semiclassicality condition \cite{Christodoulou:2023psv}, and for $ |a_\ell| \ll \lambda \delta_\ell $, one can safely drop $ \mathcal{O}({a_\ell}/{\lambda \delta_\ell} ) $, and obtain $ \mu_j\approx (2 \lambda \delta_\ell )^{M_\mathcal{C}} $.
\\
Next, by applying the decomposition \eqref{JDecomposition}, the holomorphic amplitude \eqref{final-tran-amp} can be rewritten as
\begin{eqnarray}
	W^t_\tau(H_\ell) = \mathcal N \int_{D_{g, \textbf{z}}} \mu_j ~d \mu_{g, \Omega} ~\mathcal U(g, \textbf{z}; t, H_\ell)~ e^{\lambda \Sigma(\delta_\ell,g, \textbf{z}; \textbf{k}_{\ell n})} \times e^{\lambda \delta_{\ell_f} G_\ell(g, \textbf{z}; \textbf{k}_{\ell n})} ,
	\label{holomorphic-re}
\end{eqnarray}
where
\begin{align}
	\mathcal U(g, \textbf{z}; t, H_\ell) &=  \prod_\ell\left(\sum_{a_\ell\in D^{\bar{n}}_{\omega}} e^{-a^2_\ell t + (i\gamma \zeta_\ell + F_\ell(g, \textbf{z}; \textbf{k}_{\ell n})+ G_\ell(g, \textbf{z}; \textbf{k}_{\ell n})) a_\ell}\right) , \\
	\Sigma(\delta_\ell,g, \textbf{z}; \textbf{k}_{\ell n}) &= \sum_\ell (\delta_\ell F_\ell(g, \textbf{z}; \textbf{k}_{\ell n})+i\gamma\zeta_\ell \delta_\ell).
	\label{eq:sum-delta}
\end{align}
The function $ \mathcal{U} $ exhibits continuity in both the variables $ g $ and $ \textbf{z} $, such that the generalized stationary phase theorem \cite{Hormander:2003} can be applied. The critical point equations, given by
\begin{eqnarray}
	\text{Re}\Sigma(\delta_\ell,g,\textbf{z};\textbf{k}_{\ell n}) = \delta_g \Sigma(\delta_\ell,g,\textbf{z};\textbf{k}_{\ell n}) = \delta_{\textbf{z}} \Sigma(\delta_\ell,g,\textbf{z};\textbf{k}_{\ell n}),
\end{eqnarray}
align precisely with the fixed-spin asymptotics described in \cite{Han:2011re}, allowing for direct utilization of their results in this context. Note that, there are $ 2^M $ critical points, and the data $ H_\ell $ from the semiclassical states can take a Regge-like form, leading to a geometric critical point associated with one of the three possible types of simplicial geometries. Alternatively, there might be no critical point. The data ($ \omega_\ell,\textbf{k}_{\ell n} $) is assumed to follow a Regge-like pattern, specifically chosen to exclude vector geometries.

With the stationary phase theorem in mind, the following estimate can be derived for the amplitude
\begin{equation}
	W^t_\tau(H_\ell) = \mathcal{N} \sum_c \mu_j \lambda^{M^c_{\mathcal{C}}} \mathcal{H}_c(\delta_\ell,\textbf{k}_{\ell n}) \mathcal{U}(g_c,z_c;t,H_\ell)~ e^{\lambda\Sigma(\delta_\ell,g,\textbf{z};\textbf{k}_{\ell n})} \times e^{\lambda \delta_{\ell_f} G_\ell(g,\textbf{z};\textbf{k}_{\ell n})} \left(1+\mathcal{O}(\lambda^{-1})\right) ,
	\label{amplitude-mod}
\end{equation}
where the determinant of the Hessian of $ \Sigma $ is contained in $ \mathcal{H}_c $. It is crucial to note for physical applications that in the first-order approximation, the scale $ \lambda $ only emerges as a global scaling factor $ \lambda^{M^c_{\mathcal{C}}} $ and as a linear term in the exponential. Notably, $ \mathcal{H}_c $ remains independent of $ \lambda $.

Now, it is convenient to evaluate the amplitude functions $ F_\ell(g, z; \textbf{k}_{\ell n}) $ and $ G_\ell(g, z; \textbf{k}_{\ell n}) $ in the deficit angles, as
\begin{equation}
	\begin{aligned}
		F_\ell(g, z; \textbf{k}_{\ell n}) &= -i \gamma \phi_\ell(a_{c(v)}, \delta_\ell, \textbf{k}_{\ell n}), \\
		G_\ell(g, z; \textbf{k}_{\ell n}) &= -i \gamma \alpha_\ell(a_{c(v)}, \delta_{\ell_f}, \textbf{k}_{\ell n}),
		\label{palatini-angles}
	\end{aligned}
\end{equation}
where $ \phi_\ell(a_{c(v)}, \delta_\ell, \textbf{k}_{\ell n})~ \text{and}~ \alpha_\ell(a_{c(v)}, \delta_{\ell_f}, \textbf{k}_{\ell n}) $ are the Palatini deficit angles for geometry and fermion sectors respectively. Note that the gravity part represents the dynamics of simplicial geometry. Hence, only the spins $ j_{\bar{f}} $ and spin data $ \omega_{\ell_f} $ (with $ \delta_{\ell_f} $) of the fermions are considered to be independent in the entire derivation. Following Eq. \eqref{palatini-angles}, $ \mathcal U $ can be rewritten as
\begin{eqnarray}
	\mathcal U(g_c, z_c; t, H_\ell) = \prod_\ell \left( \sum_{s_\ell \in D^{\bar{n}}_{\omega}} e^{-a^2_\ell t + i \gamma(\zeta_\ell - \phi_\ell(g, \textbf{z}; \textbf{k}_{\ell n})- \alpha_\ell(g, \textbf{z}; \textbf{k}_{\ell n})) a_\ell}\right),
	\label{eq27}
\end{eqnarray}
Since the phase $ i\gamma(\zeta_\ell-\phi_\ell-\alpha_\ell) $ is purely imaginary and does not depend on $ a_\ell $, the sum is primarily influenced by the exponential damping term $e^{a^2_\ell t}$. It is reasonable to anticipate that the rapid convergence of the sum, attributed to this damping, makes it a valid approximation to eliminate the cutoff $n$ and sum $a_\ell$ from $-\infty$ to $+\infty$ for all $\ell \in \Gamma$. This enables us to obtain an approximation as 
\begin{eqnarray}
	\sum_{a_\ell = -\infty}^{+\infty} e^{-a^2_\ell t + i \gamma (\zeta_\ell - \phi_\ell -\alpha_\ell) a_\ell} = 2\sqrt{\frac{\pi}{t}} e^{-\frac{\gamma^2}{4t}(\zeta_\ell-\phi_\ell-\alpha_\ell)^2} \vartheta_3 \bigg(-\frac{i\pi\gamma(\zeta_\ell-\phi_\ell-\alpha_\ell)}{t}, e^{-\frac{4\pi^2}{t}}\bigg) ,
\end{eqnarray}
with the third Jacobi theta function:
\begin{align}
	\vartheta_3(x, y) \equiv 1 + 2\sum_{p=1}^{\infty}y^{p^2} \cos(2px).
\end{align}
Thus, it is reasonable to anticipate
\begin{eqnarray}
	\mathcal U(g_c, z_c; t, H_\ell) \approx \prod_\ell 2\sqrt{\frac{\pi}{t}} e^{-\frac{\gamma^2}{4t}(\zeta_\ell-\phi_\ell-\alpha_\ell)^2} \vartheta_3\left(-\frac{i\pi\gamma(\zeta_\ell-\phi_\ell-\alpha_\ell)}{t}, e^{-\frac{4\pi^2}{t}}\right).
\end{eqnarray}
Finally, substituting every modification in hand to Eq. \eqref{holomorphic-re} (or Eq. \eqref{amplitude-mod}), one gets
\begin{align}
	W^t_\tau(H_\ell) = ~ &\mathcal{N} \sum_c \lambda^M \mu(a) \prod_\ell \left[ e^{-\frac{\gamma^2}{4t}(\zeta_\ell-\phi_\ell-\alpha_\ell)^2} \vartheta_3\left(-\frac{i\pi \gamma(\zeta_\ell-\phi_\ell-\alpha_\ell)}{t}, e^{-\frac{4\pi^2}{t}}\right) \right]
	\nonumber \\
	&\times
	e^{ \Sigma_\ell \left[-i\gamma \lambda \delta_\ell \phi_\ell(a_{c(v)},\delta_\ell,\textbf{k}_{\ell n}) - i\gamma \lambda \delta_{\ell_f} \alpha_\ell(a_{c(v)},\delta_{\ell_f},\textbf{k}_{\ell n}) +i\gamma \zeta_\ell \lambda \delta_\ell \right] } \, .
\end{align}
The exponent $ M $ typically takes on a half-integer value determined by both the rank of the Hessian matrix at the critical point and the combinatorial properties of the 2-complex $ \mathcal{C} $. The function $ \mu(a) $ encompasses the summation measure over the spins and the evaluation of the Hessian at the critical point.

For the present work, the theta function can be approximated safely as $\vartheta_3 \approx 1$ (for any details on the consideration, follow \cite{Christodoulou:2023psv}), such that
\begin{equation}
	W^t_\tau(H_\ell) \approx \mathcal{N} \sum_c \lambda^M \mu(a) \prod_\ell e^{\left[-\frac{\gamma^2}{4t}(\zeta_\ell-\phi_\ell-\alpha_\ell)^2 +i\gamma(\zeta_\ell-\phi_\ell)\omega_\ell - i\gamma \alpha_\ell \omega_{\ell_f}\right]} \left(1+\mathcal{O}(\lambda^{-1})\right) \, ,
	\label{amplitude-estimate}
\end{equation}
which can be generalized to accommodate all the geometrical arguments of critical points.

Note that the spins \( j_{\ell} \) are centered around the area data \( \omega_{\ell} \), corresponding to the triangle areas \( A_{\ell} = \omega_{\ell} \, \hbar \) within a triangulation of \( \mathcal{B} \). Consider a triangulation where all discrete areas scale with \( m^2 \), the natural area scale of the spacetime, expressed as
\(
A_{\ell} = m^2 \, \hbar \, \delta_{\ell}
\),
where the spin data \( \delta_{\ell} \) are of order unity. Although the spin data \( \delta_{\ell} \) can vary with \( T/m \), akin to the boundary data in \cite{Christodoulou:2016vny}, the area data \( \omega_{\ell} \) takes the form
\(
\omega_{\ell}(m,T) = \lambda \; \delta_{\ell}(X)
\),
where \( \delta_{\ell}(X) \) are numbers of order unity for all \( X \) \cite{Christodoulou:2018ryl}. Here, \( \lambda \equiv \frac{m^2}{\hbar} \) and \( X \equiv \frac{T}{m} \).

Referring to \cite{Christodoulou:2018ryl}, note that all proper areas in the HR spacetime are of the form \( m^2 \, \delta(X) \) with \( \delta(X) \) as a function of \( X \), also supported by dimensional analysis. The areas \( A_{\ell} \) result from classical discretization, where \( \hbar \) appears only as a constant relating to unit choice. Within geometrical units (\( G=c=1 \)), where length, time, and mass are all of dimension \( \sqrt{\hbar} \), the embedding data \( \zeta_{\ell} \) and the 3d normal data \( \textbf{k}_{\ell n} \) are solely functions of \( X \).

The semiclassicality parameter \( t \) regulates the coherence properties of states. As depicted in \eqref{hk-state}, it must be small and positive. Following \cite{Bianchi:2009ky,Thiemann:2000ca}, \( t \) corresponds to a dimensionless physical scale, thus proportionate to a positive power of \( \hbar \).

The area data \( \omega_\ell \) and 3d normal data \( \textbf{k}_{\ell n} \) are assumed to exhibit Regge-like characteristics \cite{Barrett:2009gg}. This assumption entails that \( \omega_\ell \) and \( \textbf{k}_{\ell n} \) describe a piecewise flat geometry for the 4d simplicial triangulation that is dual to the 2-complex \( \mathcal{C} \). Importantly, this assumption does not pertain to the embedding data \( \zeta_\ell \). It indicates the existence of a critical point for the partial amplitude given by Equation \eqref{final-tran-amp}, corresponding to a classical discrete \emph{intrinsic} geometry. The intrinsic geometry specified by \( \omega_\ell \) and \( \textbf{k}_{\ell n} \) could be Lorentzian, 4d Euclidean, or degenerate. The degenerate scenario corresponds to 4-simplices with vanishing four-volume.

Now, the simplification of Eq. \eqref{amplitude-estimate} takes the form
\begin{eqnarray}
	W(\omega_\ell, \zeta_\ell, \textbf{k}_{\ell n}, t)\approx \mathcal{N}\sum_c \lambda^M\mu(\delta)\prod_\ell e^{-\frac{{\Delta_\ell}^2}{4t}+\Delta_A}\left(1+\mathcal{O}(\lambda^{-1})\right ),
	\label{final}
\end{eqnarray}
where $\Delta_\ell \equiv \gamma \zeta_\ell-\beta(\phi_\ell+\alpha_\ell)+\Pi_\ell$, and $\Delta_A \equiv i\gamma((\zeta_\ell-\phi_\ell)\omega_\ell-\alpha_\ell \omega_{\ell_f})$, for given spin data $\delta_\ell$, 3d normal data $\textbf{k}_{\ell n}$ and embedding data $\zeta_\ell$. The above expression is the final estimation for the transition amplitude of fermions from the stationary phase approximation in $ \lambda $, resulting from the appropriate manipulation of Eq. \eqref{final-tran-amp}.

Few important points to note here is that the contribution from $\Pi_\ell$ introduces an additional phase factor in the Lorentzian intertwiners, as discussed in \cite{Barrett:2009mw,Bianchi:2011hp}. The exponent $ M $ typically takes on a half-integer value, which relies on factors such as the rank of the Hessian at the critical point and the combinatorial properties of the 2-complex $ \mathcal{C} $. The function $ \mu(a) $ incorporates both the summation measure over spins and the evaluation of the Hessian at the critical point. Notably, neither the summation measure nor the Hessian varies with $ \lambda $.

The estimation \eqref{final} holds true for all three possible types of geometrical critical points. When $ \omega_{\ell} $ and $ \textbf{k}_{\ell n} $ denotes a Lorentzian geometry, $ \beta=\gamma $, such that, $ \Pi_\ell= \{ 0~ \text{for~thick~wedge},~ \pi~ \text{for~thin~wedge}\} $. Additionally, for a degenerate 3d geometry, i.e., $ \beta=0 $, the dihedral angle $ \phi_\ell(\omega_{\ell},\textbf{k}_{\ell n})= \{ 0~ \text{for~thick~wedge},~ \pi~ \text{for~thin~wedge}\} $. Consequently, when $ \omega_{\ell} $ and $ \textbf{k}_{\ell n} $ describe a 4d Euclidean geometry, such that $ \beta=1 $ and $ \Pi_\ell=0 $; $ \phi_\ell(a_{c(v)}, \delta_{\ell}, \textbf{k}_{\ell n} ) $ and $ \alpha_\ell(a_{c(v)}, \delta_{\ell_f}, \textbf{k}_{\ell n} ) $ represent Palatini deficit angles \cite{Christodoulou:2018ryl,Christodoulou:2023psv}.

Notice, the geometry and wedge parameters $ \beta $ and $ \Pi_\ell $ did not appear till Eq. \eqref{final} in the estimation, since the derivation has been performed for Palatini deficit angles, as in Eq. \eqref{palatini-angles}. It is therefore expected for future works to include these parameters in the estimation. However, readers will see that this would not affect the final result of the study.

Each critical point of Eq. \eqref{final-tran-amp} possesses a degeneracy of $2^M$, representing the various configurations of the orientation $a(v)$ of the tetrad, where $a(v)$ can take values of $+1$ or $-1$ at each vertex of $\mathcal{C}$. All $2^M$ critical points corresponding to a given $\delta_\ell$ and $\textbf{k}_{\ell n}$ yield the same intrinsic Regge geometry. This results in a summation over the configurations of $a(v)$ in the estimate \eqref{final} \cite{Rovelli:2012yy,Christodoulou:2012sm,Immirzi:2016nnz,Vojinovic:2013faa}. This summation reflects the origin of such models from tetradic actions like the Palatini and Holst actions in General Relativity. The Palatini deficit angles $\phi_\ell(\delta_\ell)$ and $\alpha_\ell(\delta_{\ell_f})$ also depends on $a(v)$ and corresponds to the standard Regge deficit angle when $a(v)$ is uniform, i.e., when $a(v)=1$ for all vertices of the 2-complex $\mathcal{C}$ or $a(v)=-1$ for all vertices of $\mathcal{C}$. 

However, note that the summation over orientation configurations \( a(v) \) can be disregarded \cite{Christodoulou:2018ryl}, and the complete amplitude is effectively approximated by retaining solely the contribution from the dominant co-frame configuration. A similar rationale in an alternate context was presented in \cite{Bianchi:2008ae}. It is worth mentioning that instead of the EPRL model, one may employ the ``proper vertex'' model \cite{Engle:2012yg,ChaharsoughShirazi:2015uuu,Engle:2011un}, where only a single co-frame orientation configuration remains in Eq. \eqref{final}, corresponding to the Regge case where \( a(v) = 1 \) at each vertex.

%

Now, as we have arrived at the final estimation, we need to wrap up a few things very quickly. The transition amplitude is influenced by the bounce time $T$ exclusively through $X$ only, such that $ W(\omega_\ell, \zeta_\ell, \textbf{k}_{\ell n}, t)=W(m,X)$, where in particular, $m$ is present through $\lambda$ and $t$. Following the above arguments, Eq. \eqref{final} is approximated using the dominant co-frame orientation, as
\begin{equation} \label{eq:squaredAmpEstim}
	\vert W \vert^2 \approx  \lambda^{2M} \mu(\delta)^2 e^{-\frac{\sum_\ell \Delta_\ell^2 }{2t}} \left( 1 + \mathcal{O}(\lambda^{-1}) \; \right).
\end{equation}
The amplitude diminishes exponentially as $\hbar \rightarrow 0$, as anticipated for a tunneling phenomenon, unless all embedding discrepancies $\Delta_\ell$ vanish. This scenario is not plausible since it would imply the existence of an exact classical solution of the discretized theory, connecting a black hole in the past to a white hole in the future. However, referring to the discussions, the crossing time of Eq. \eqref{cross-time} is arrived at
\begin{equation} 
	T_c = m \frac{\int d X \; X \,  \mu(X) \; e^{-\frac{1}{2t} \sum_\ell \Delta_\ell^2(X) }}
	{\int d X \; \mu(X) \; e^{-\frac{1}{2t}\sum_\ell \Delta_\ell^2(X)}},
	\label{cross-time_final}
\end{equation}
with the upper limit $T_c=mf(\gamma,t)$, where the function of semiclassicality parameter $t$ and $\gamma$, i.e. $f(\gamma,t)$, can be well approximated by the details of discretization. Still, if a minimum is present in $\sum_\ell \Delta^2_\ell(X)$, for some $X=X_0$, $T_c$ must be independent of such details. Therefore, assuming such a minimum to be present, 
\begin{equation}
	T_c=mX_0(\gamma)(1+\mathcal{O}(t)),
\end{equation}
which is the final form of the desired crossing time, found to be directly dependent on $ \Delta_\ell $. Referring to \cite{Christodoulou:2018ryl}, the crossing time of black hole is estimated by using $(\Delta_\ell)_{\text{BH}}=\gamma\zeta_\ell-\phi_\ell$, whereas, for fermions, it is determined to be $(\Delta_\ell)_{\text{FERMION}}=\gamma\zeta_\ell-(\phi_\ell+\alpha_\ell)$. 


\section{Results and Discussions}

The hypotheses coming out of the final discussion have major significance on the black-to-white hole bounce model. The black hole tunnel to white hole through the quantum tunneling region near the trapping horizon, and so do the external fermions. The aim is, therefore, remains to determine fermion's characteristic time scale in a physically pertinent theory. On the mathematical standpoint, the derivations are quite well-known, however, in the physical context, the results are significantly intriguing. 

From the definition of 4d Euclidean deficit angle:
\begin{equation}
	\phi_f(l)= 2\pi - \sum_{v\in f} \Theta_{vf}(\ell)
	\nonumber
\end{equation}
where $ \Theta_{vf}(\ell) $ are the dihedral angle between two adjacent tetrahedra in the 4-simplex $ v $, having face $ f $. In the Euclidean case, these dihedral angles are always positive. Consequently, when $ \sum_{v\in f} \Theta_{vf}(\ell) =2\pi $, the deficit angle vanishes, resulting in flat local curvature. If it exceeds $ 2\pi $, the deficit angle becomes negative, leading to pinched local curvature (negative curvature). Conversely, if it is less than $ 2\pi $, the deficit angle is positive, indicating stretched local curvature (positive curvature).
\\
Therefore, for a flat curvature, the tunneling time of the black hole and fermions are equal, i.e., the geometry and exterior particles emerge to the white hole together. But, when it is a negative local curvature, $ \Delta_{\text{BH}} < \Delta_{\text{FERMION}} $, leading to the fact that, excess fermions may accumulate in the black hole exterior when the black hole has already been tunneled through. Alternatively, for positive local curvature, $ \Delta_{\text{BH}} > \Delta_{\text{FERMION}} $. Therefore, fermions consume a shorter time to reach the white hole exterior. This may open up a window for fermions to interact with the future state of a black hole. The idealized scenario of the local positive curvature is significantly equivalent to the Einstein-Rosen \textit{``bridge''} \cite{Einstein:1935tc}, where they hypothesized the construction of two asymptotic geometries glued at $r=2m$. However, the similarity of its physical context with the model, still remains a question, which will be investigated further in a subsequent work.

The study thus revisits numerous unresolved issues, leaving many of them still open, while also bringing forth several new questions. The model could also emerge as a useful tool for better understanding the origin of the information paradox and Hawking radiation, using the tunneling characteristics.


\end{document}